\documentclass[epj,referee]{svjour}

\def\EQ{\begin{equation}}
\def\EN{\end{equation}}
\def\EQA{\begin{eqnarray}}
\def\ENA{\end{eqnarray}}

\usepackage{amsmath}
\usepackage[dvips]{graphicx}

\begin{document}

\title{Stability of a nonlinear  oscillator with random  damping}
\author{N. Leprovost\inst{1} \and S. Auma\^{\i}tre\inst{1}
  \and K. Mallick\inst{2}}
\institute{Laboratoire de Physique statistique de l'ENS,
 24 rue Lhomond, 75231 Parix cedex 05, France \\
  \email{nicolas.leprovost@lps.ens.fr}
\and Service de Physique th\'eorique, CEA Saclay, 
 F-91191 Gif sur Yvette Cedex, France}

\date{Received: date / Revised version: date}

\abstract{A noisy  damping parameter in the equation of motion of a
 nonlinear  oscillator renders the fixed point  of the  system
 unstable when the amplitude of the noise is sufficiently large.
 However, the stability diagram of the system can not be predicted
 from the analysis  of the moments  of the linearized equation.  
  In the case of a white noise, an   exact formula for 
  the Lyapunov exponent of the system  is derived.
  We then  calculate  the critical
 damping for which  the  {\em nonlinear} system  becomes unstable.
 We also characterize the  intermittent  structure of the
 bifurcated  state above  threshold and  address the effect of
 temporal correlations of the noise by considering  an
 Ornstein-Uhlenbeck noise.}


\maketitle

\section{Introduction}
It is a well-known fact  that a multiplicative noise acting on a
dynamical system can generate unexpected  phenomena  such as
stabilization  \cite{Graham82b,Lucke85}, stochastic  transitions
 and  patterns \cite{VandenBroeck94} or stochastic resonance
\cite{Gammaitoni98}.  As far as the stability of a random dynamical
system is concerned,  the exactly solvable example of a nonlinear first-order
Langevin equation shows that the behaviour of the moments of the
linearized equation can be misleading~: higher moments seem to be
always unstable although it is known that the 
critical value of the control  parameter is the same as  that of the
deterministic system \cite{Graham82}.  This apparent contradiction is
due to the existence of long tails in the stationary probability
distribution of the linearized equation which are suppressed when the
 nonlinearities are  taken into account.  The same type of
conclusion can be drawn for a second-order  system, namely a
nonlinear oscillator with fluctuating frequency. The energetic
stability analysis of the linearized system has been performed long ago
\cite{Bourret71,Bourret73} but it leads  to an erroneous phase diagram.
In fact, because of the nonlinearities, the noise can even enhance  the
stable phase \cite{Graham82b,Lucke85}. Besides, a  non-perturbative reentrant
transition can occur~: a noise of small amplitude stabilizes the system,  but a
strong noise  destabilizes it.  Similar features  hold
in spatially extended systems such as the Ginzburg-Landau
and Swift-Hohenberg equations driven by a noisy control parameter
\cite{VandenBroeck94,Becker94}. 
  Here again,  the observed threshold shift can not be
tackled by the  analysis of the moments' behaviour alone.

 In this paper, we study the  nonlinear  oscillator with
 random damping. This  dynamical system, in which
 the noise acts on the velocity variable rather
 than on the position, 
 provides another example  of  a bifurcation  induced by a 
 multiplicative noise.  This model  was  introduced in the study of
 the generation of water waves by wind \cite{West81} where the
 turbulent fluctuations in the air flow is modeled by a noise. This
 model is also relevant in  the study of dynamical systems  with an
 advective term where the corresponding velocity fluctuates.
  This problem has been investigated in
 \cite{VKampen74,Gitterman04} via  the stability of the moments  of
 the {\sl linearized} equation. Here, we demonstrate that,  as in the
 case of the random frequency oscillator,  the moments stability
 analysis has no relevance and can not be used to derive
  the instability threshold.
  We shall obtain  the exact phase diagram of the system  thanks to an
 instability criterion valid for general nonlinear stochastic
 dynamical systems  \cite{arnold,Mallick03}.

 The outline of this work is  as follows.  In section
 \ref{sec:model}, we define  the model and summarize the results obtained from the stability analysis of the
 moments.  The main part of the paper (section \ref{Lyapunov}) is
 devoted to the calculation of  the Lyapunov exponent of the system from
 which the  marginal stability curve is  drawn.  We then  analyze the
 nature of the bifurcation and show that the bifurcated state is
 intermittent (section \ref{Intermittency}).  In the last section, we
  consider an Ornstein-Uhlenbeck noise to investigate the effect
 of temporal correlations on the instability threshold.

\section{Definition of the  model and basic results}
\label{sec:model}
 We  consider  the following stochastic dynamical system: \EQ
\label{model1}
\ddot{x} + [r + a x^2-\xi(t)] \dot{x} + \alpha x = - b x^3 \; , \EN
where $\xi(t)$ is a white noise with intensity D  and the parameters
$a$ and $b$ are taken to be positive in order to  have a stabilizing
effect. The mean value of the damping $r$ is also positive whereas the
coefficient $\alpha$  of the linear term  can
be either positive or negative.  In fact, as far as the stability of
the fixed point is concerned,  the  $\alpha < 0$ case  can be reduced to
the $\alpha > 0$ case as  follows. For  $\alpha < 0$ and in the absence of
noise, the oscillation around the $x=0$ position is unstable and the
new equilibrium position in the saturated regime is $x_0 =\pm
\sqrt{-\alpha/b}$.  Defining  $y = x - x_0$, we observe that  $y$
satisfies:  \EQA \ddot{y} &+& (r - \frac{a \alpha}{b} + 2 a x_0 y + a
y^2) \dot{y} - 2 \alpha y \\ \nonumber &=& \dot{x} \xi(t) - b (3x_0
y^2 - y^3) \; . \ENA  The linear part of this equation is the same as
that of equation (\ref{model1}) once the substitutions $r'=r -
a\alpha/b$ and $\alpha'=-2\alpha$ are made.  The coefficient
$\alpha'$ is now  positive. We  thus take  $\alpha > 0$.
Before proceeding, we write  time   and   position in dimensionless
 units by multiplying $t$ and 
$x$  by the factors $1/\sqrt{\alpha}$ and $\sqrt{a/r}$, respectively.
 Defining  the following parameters: \EQ \gamma = \frac{r}{\sqrt{\alpha}}
\quad , \qquad \Delta = \frac{D}{\sqrt{\alpha}} \quad \text{and}
\qquad k = \frac{b r}{a \alpha} \; , \EN  
equation (\ref{model1}) becomes 
  \EQ
\label{model2}
\ddot{x} + \gamma (1 + x^2) \dot{x} + x = \dot{x} \xi(t) - k x^3 \; ,
\EN 
 where  the autocorrelation of the white noise  is given by $\langle
\xi(t) \xi(t') \rangle = \Delta \delta(t-t')$  (as usual, the brackets
represent an average on the realizations  of the noise).  The aim of
this work is to study the  stability of the  fixed  point $x=0$ of
equation~(\ref{model2}) as a function of the values  of the different
parameters $\gamma$, $\Delta$ and $k$.

We now summarize the results for the stability of the moments of the
 amplitude $x$, obtained by {\sl linearizing} equation~(\ref{model2})
 around the fixed point $x =0$.  It can be  shown that the first
 moment satisfies~:
  \EQ \frac{d^2}{d \, t^2} \langle x \rangle + (\gamma -
 \frac{\Delta}{2}) \frac{d}{d \, t} \langle x \rangle + \langle x
 \rangle = 0 \; . \EN  The first moment remains bounded  as $ t \to
 \infty$ provided  $\gamma > \Delta/2$. The effect  of the noise
 is therefore to enhance the unstable phase. An  intuitive 
 argument of  van Kampen
 \cite{VKampen74} explains why this should be the case~: positive and
 negative fluctuations are equiprobable, but because the noise is
 multiplied by the velocity, the negative fluctuations have a stronger
 effect because  they tend to increase the velocity  (whereas the
 positive fluctuations  decrease the velocity  and  have a lesser
 impact). For the second order moment, one must consider simultaneously
  $\langle x^2 \rangle$, $\langle \dot{x}^2 \rangle$ and
 $\langle x\dot{x} \rangle$,  and study the linear system that
 couples them. 
 These three quantities  become zero in the
 long  time limit provided  $\gamma > \Delta$. In this  case,
  the system is  said to be `energetically' stable.
 The instability threshold  thus depends on  the  moment  which is
 considered,  due to  the interplay between noise and
 nonlinearities.  Therefore, in contrast to the deterministic case, a
 naive linear stability ana\-lysis fails to  lead to conclusive results.
 A more refined criterion is needed  to determine  stability.

\section{The Lyapunov exponent}
\label{Lyapunov}
The correct criterion  that determines the  stability of the 
 fixed point $x=0$ is based upon  the Lyapunov exponent $\Lambda$, defined
 as 
 \EQ 
 \Lambda =  \frac{1}{2} \langle  \, \partial_t \ln [{\delta x}^2] \, 
 \rangle = \langle \frac{\partial_t({\delta x})}{x} \, \rangle \, ,
\label{eq:defLyap}
 \EN 
 where ${\delta x}$ satisfies equation~(\ref{model2}) linearized
 around  $x=0$.  It has been shown \cite{arnold,Mallick03,Hansel89} 
  that the sign of the 
 Lyapunov exponent,  calculated with the {\sl linear part of the
 equation},  monitors the instability of the nonlinear oscillating
 system.    When $\Lambda$ is negative,
 the trajectories of the  nonlinear  system (\ref{model2}) almost
 surely decay to zero and in the long time limit, the oscillator becomes
  localized in   its rest position $x=0$.  On the contrary, if
 $\Lambda > 0$, the fixed point  $x=0$ is 
 unstable and the stationary probability
 density of the oscillator is extended.

 We now calculate the Lyapunov exponent   as a function of the
parameters  $\gamma$ and  $\Delta$. Defining the variable $z$ as
$z=\partial_t({\delta x})/{\delta x}$,  the 
equation  satisfied  by   $z$ is given by 
\EQ
 \label{zpoint}
\dot{z} = - 1 - \gamma z - z^2 + z \xi(t) \; . \EN 
 This  is a Langevin
 equation  involving  the variable $z$ alone and  coupled to  a
 multiplicative noise (to be interpreted in the Stratonovich sense).
  By definition~(\ref{eq:defLyap}), we observe that 
  \EQ
\Lambda(\gamma,\Delta) = \langle z \rangle \, .
 \label{Lyapz}
 \EN 
   Using standard techniques \cite{VKampen81,Gardiner84}, we obtain the
 Fokker-Planck equation for the probability density  $P(z,t)$ at time
 $t$.  The stationary probability density $P_s(z)$ satisfies: 
 \EQ
\label{FPstat}
\frac{\Delta}{2} \partial_z\bigl[ z P_s(z)\bigr] + \frac{1+\gamma z +
z^2}{z} P_s(z) = \frac{J}{z} \; , \EN where $J$ is the constant
current of probability. Here, the current $J$ does not vanish;  
intuitively, this is due to the fact
that $z$ plays the role of an  angular variable in phase space and
 must be interpreted  as a compact variable  (see \cite{Risken}
for more details).  Equation~(\ref{FPstat})  can be solved by 
variation of constants method  and we obtain 
 \EQA \nonumber P_s(z) = \frac{2J}{\Delta}
&\int_c^z& \vert \frac{t}{z} \vert^{1+2\gamma/\Delta} \frac{1}{t^2}
\exp\Bigl[-\frac{2}{\Delta}\bigl(\Phi(z) - \Phi(t)\bigr)\Bigr] \, dt
\\ &+& A \vert z \vert^{-1-2\gamma/\Delta} \exp\Bigl[-\frac{2
\Phi(z)}{\Delta}\Bigr]\; ,
\label{Pstat1}
\ENA where $\Phi(t) = t - 1/t$;  the constant  $A$
 and the  reference  point  $c$  
 are  not determined at this stage.  We must impose $A=0$
 so  that $P_s(z)$  is
 integrable  at infinity  (this implies  $J \neq 0$).
  Furthermore,  the function
$\exp[\Phi(t)]$ is exponentially divergent  when  $t \to 0^{-}$;
this remark  fixes the choice of the  value  of $c$
 in  order to ensure that $P_s(z)$ is integrable  at the origin 
   (the cases $z < 0$ and $z > 0$ 
 must be analyzed separately  and   two different reference points
 must be chosen).
  Finally, we obtain,  \EQ
\label{Pstat2}
P_s(z) =
\begin{cases}
\frac{1}{N} \int_{-\infty}^z H(z,t) \, dt & \text{if} \quad z <0 \\
\frac{1}{N} \int_{0}^z \; H(z,t) \, dt & \text{if} \quad z > 0 \\
\end{cases}
\EN where $H(z,t)$ represents the function under the integral sign in
(\ref{Pstat1}).  Using equation (\ref{FPstat}), we find that  the
probability density (\ref{Pstat2}) satisfies  \EQA P_s(z) &\sim&
\frac{J}{z^2} \quad  \, \,  \text{for} \,\,   z \longrightarrow \pm
\infty   \label{Psinf}\, ,         \\  P_s(z) &\sim&  J \qquad
\text{for}\,\,  z \longrightarrow 0
 \label{Ps0} \, .
\ENA  This quadratic decay insures that the probability density is
 indeed normalizable at infinity.
%%%%% In appendix \ref{AppendixA}, we examine in detail 
%%%%% the behaviour of $P_s(z)$ near the origin and near infinity. 
 The normalization constant  $N$ can calculated exactly
 \cite{gradstein} and we obtain \EQ  N= \pi^2
 \Bigl[J_{2\gamma/\Delta}^2\bigl(\frac{4}{\Delta}\bigr) +
 Y_{2\gamma/\Delta}^2\bigl(\frac{4}{\Delta}\bigr) \Bigl] \; ,
 \label{eq:N} \EN where
$J$ and $Y$ are  Bessel functions of the first kind and
second kind, respectively. This formula was  derived 
 in \cite{Bouchaud87} for the study of  diffusion in a random
medium.
 
 According to equation~(\ref{Lyapz}), the  Lyapunov exponent is equal
  to the mean value of $z$ calculated  with respect to $P_s(z)$; this quantity
  will be evaluated in the sense of principal parts
  \EQ  
 \Lambda =
  \lim_{M\rightarrow \infty} \int_{-M}^{+M} z P_s(z) dz \; .  
 \EN
  (Thanks to equation~(\ref{Psinf}), the logarithmic 
  divergencies at  $-\infty$ and $+\infty$ cancel with each other by
  parity).  An exact closed formula for the  Lyapunov exponent 
  can then be derived   and is given by 
 \EQ
\label{Lyapunovtheo}
\Lambda = \frac{8}{N} \int_0^{+\infty} K_1\Bigl(\frac{8 \sinh
x}{\Delta}\Bigr)
\sinh\Bigl[\bigl(1-\frac{4\gamma}{\Delta}\bigr)x\Bigr] \; dx \, , \EN
 where  $K_1$ is  a modified Bessel function of the second
  kind and the normalization constant $N$  is 
 given in equation~(\ref{eq:N}).
From this expression,  the instability threshold is readily found.
Indeed,  because the Bessel function is  always positive, the sign of the
Lyapunov is the same as that of the argument of the hyperbolic sine.
 We conclude that  the fixed point $x =0$ is stable when  \EQ \gamma >
\frac{\Delta}{4}  \; .  \EN 
 We find, as usual, that the stable phase
 is wider than what  moments stability predicts.
  Reverting to the initial variables, we observe that the
instability threshold is given by $r =  D/4$ which  does not depend on
the  `linear frequency'  $\alpha$ of the oscillator. In the $(r, D)$
plane, the critical curve in the stability diagram of the $x=0$
 fixed point is simply a straight line. We emphasize that this exact 
 result is much simpler than  that  obtained for the Duffing
 oscillator with random frequency \cite{Mallick03}.

\begin{center}
\begin{figure}
\includegraphics[scale=0.4,clip]{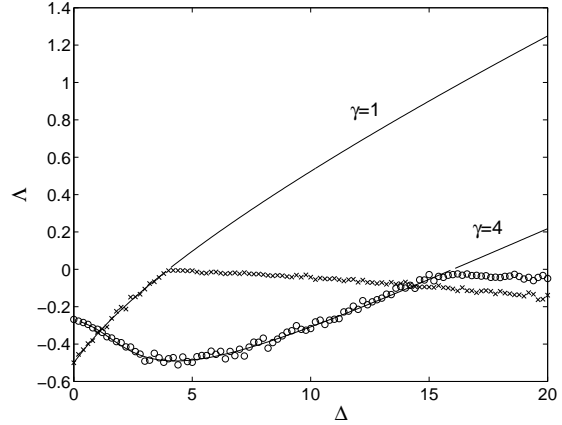}
\caption{\label{Lyapgamma1} Typical evolution of the Lyapunov exponent
 as a function of  the noise amplitude  $\Delta$. The
 circles/crosses are obtained by   numerical simulations
 of the nonlinear equation in  the stationary regime
  for  $\gamma=4$ and $\gamma=1$.  The solid lines
 represent  the theoretical values  (\ref{Lyapunovtheo}) for the
 linearized equation.}
\end{figure}
\end{center}

 In figure \ref{Lyapgamma1}, we  plot  the Lyapunov exponent of the
  nonlinear equation~(\ref{model2}) for $\gamma = 1$ and $\gamma=4$.
  We observe that,  for  $\Lambda < 0$,  the numerical findings and the
  analytical prediction are in perfect agreement.  For  higher values
  of the noise, the nonlinear term can not be neglected~: 
  whereas the linear Lyapunov exponent, given by
  equation~(\ref{Lyapunovtheo}) continues to grow,  the Lyapunov
  exponent of the nonlinear equation saturates to a value close to
  0. This indicates that the nonlinear  system has gone through a
  bifurcation~:  the fixed point $x =0$  has become unstable 
  and in the long time limit the system reaches 
  an extended  stationary state 
  (in our case a noisy limit cycle). A remarkable
  feature  of the curves displayed in figure \ref{Lyapgamma1}, is that
  noise can have a stabilizing effect at small amplitudes~: we notice that
  the curve for $\gamma = 4$  decreases for small values of $\Delta$
  and then increases.   A  closer inspection shows that  this
  non-monotonic behaviour occurs  when $\gamma >2$. This feature seems
  to contradict the argument, presented in section \ref{sec:model},
  stating that a noise in the damping term always has a distabilizing
  effect \cite{VKampen74}.  This contradiction stems from the
  asymmetry of  growth rate of the deterministic system around $\gamma
  = 2$. Indeed, for $\gamma < 2$ the growth rate is a linear function
  of $\gamma$ ($\sigma = - \gamma / 2$),  whereas  it has a very
  different expression for $\gamma > 2$ ($\sigma = - \gamma / 2 +
  \sqrt{\gamma^2-4}/2$).  The  positive and negative
  fluctuations of  the growth rate   are 
  therefore intrinsically  asymmetric,  
   in contrast to  the implicit  assumption   in  van  Kampen's argument.

\section{Intermittency above the threshold}
\label{Intermittency}
\begin{center}
\begin{figure}
\includegraphics[scale=0.23,clip]{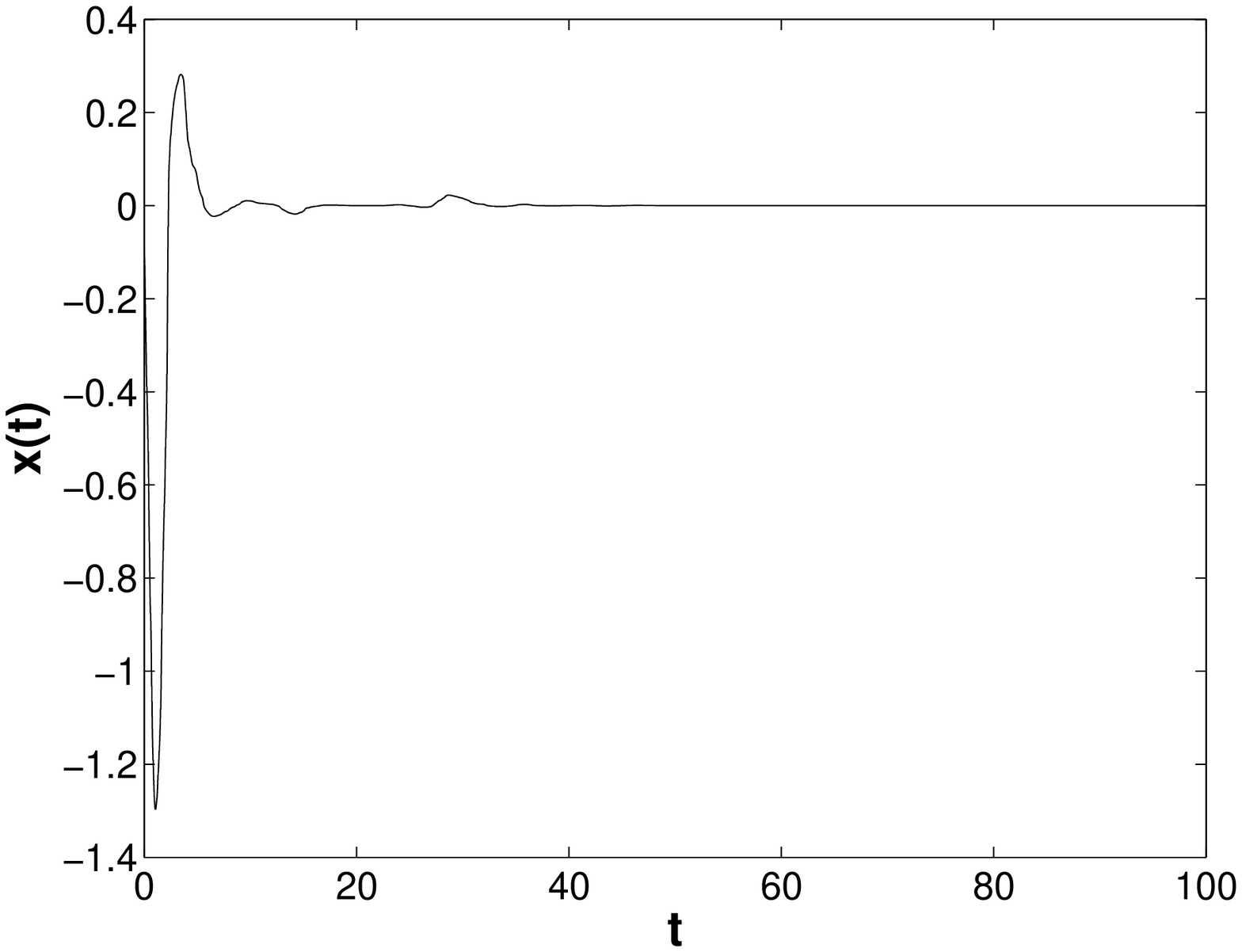}
\includegraphics[scale=0.23,clip]{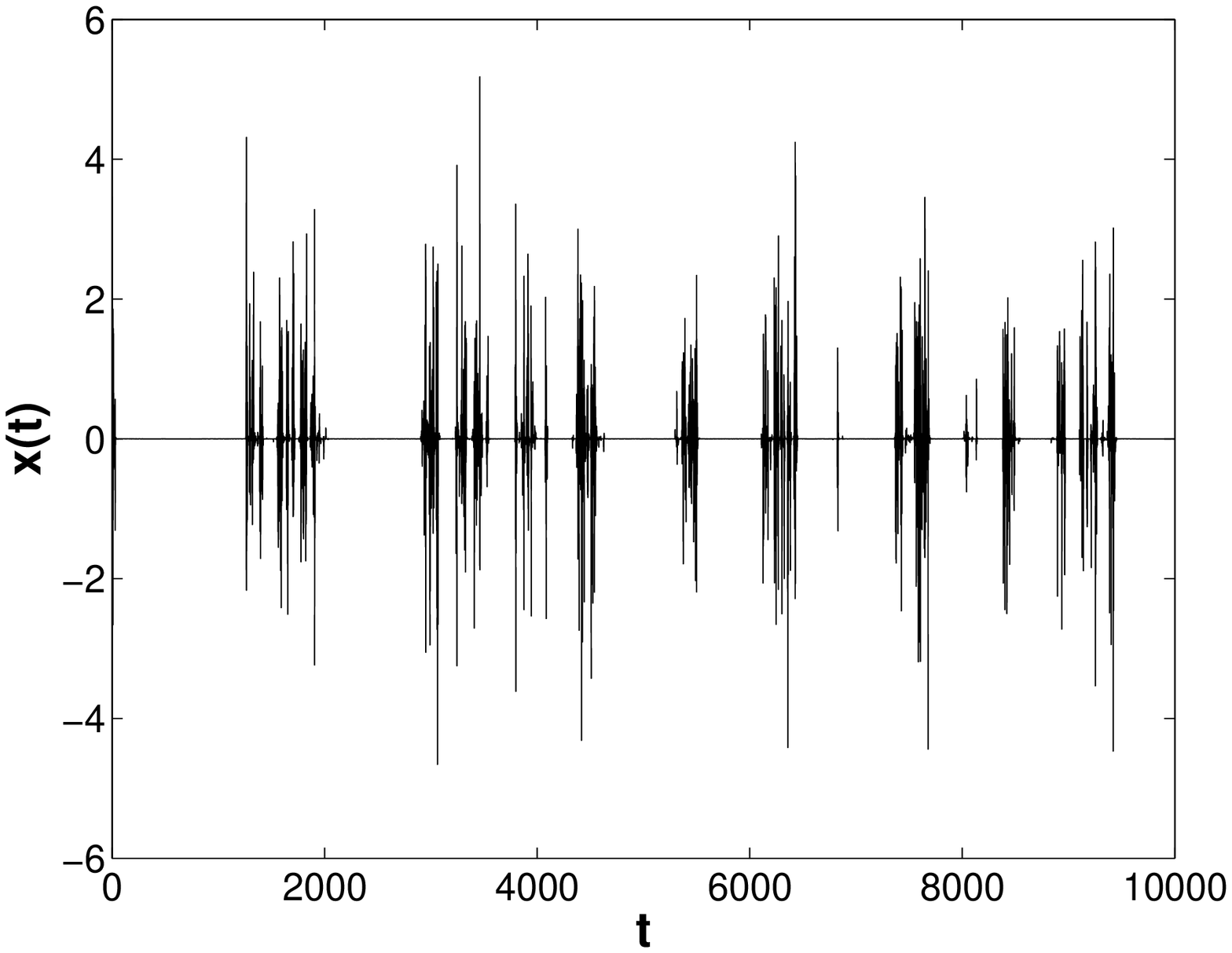}
\includegraphics[scale=0.23,clip]{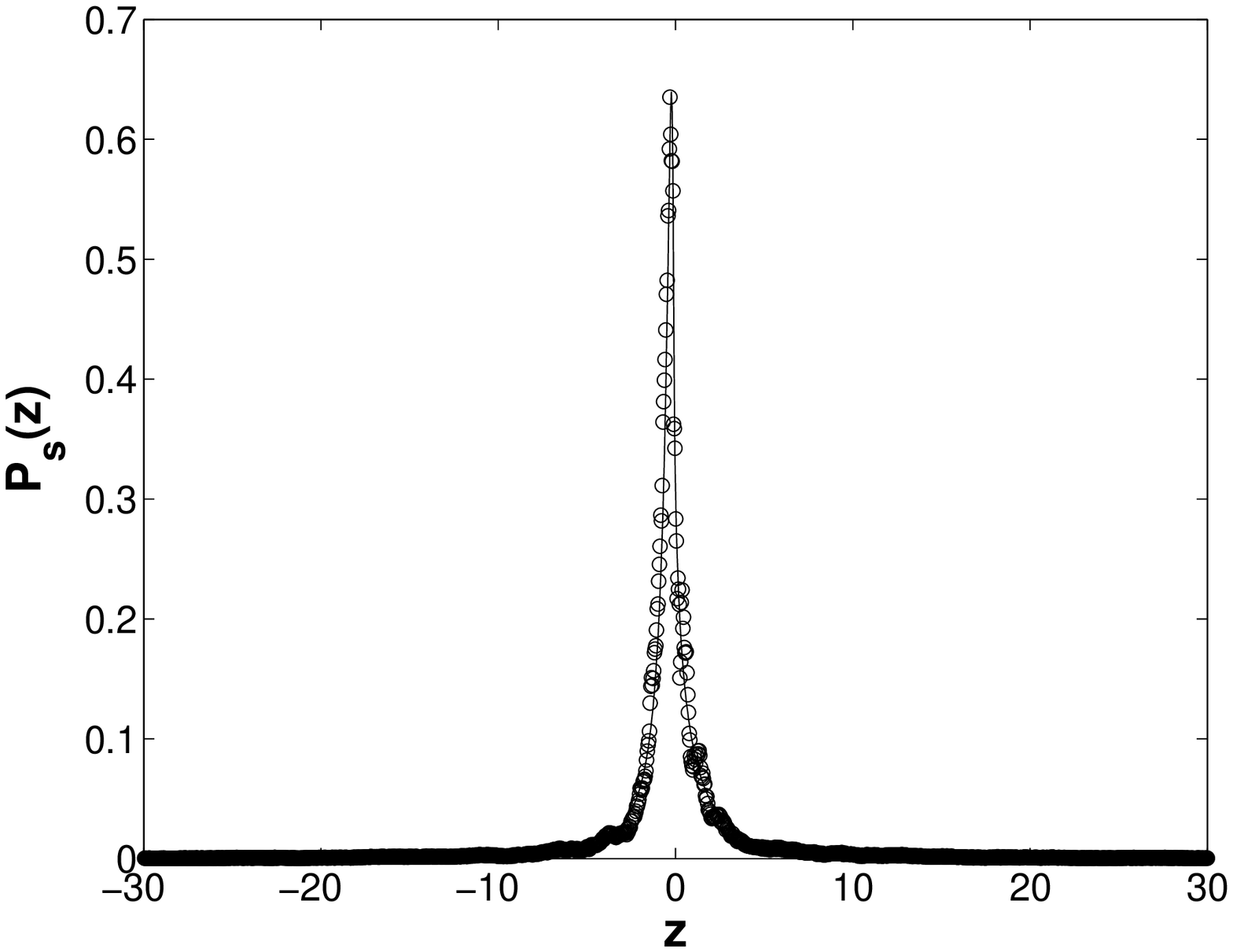}
\includegraphics[scale=0.23,clip]{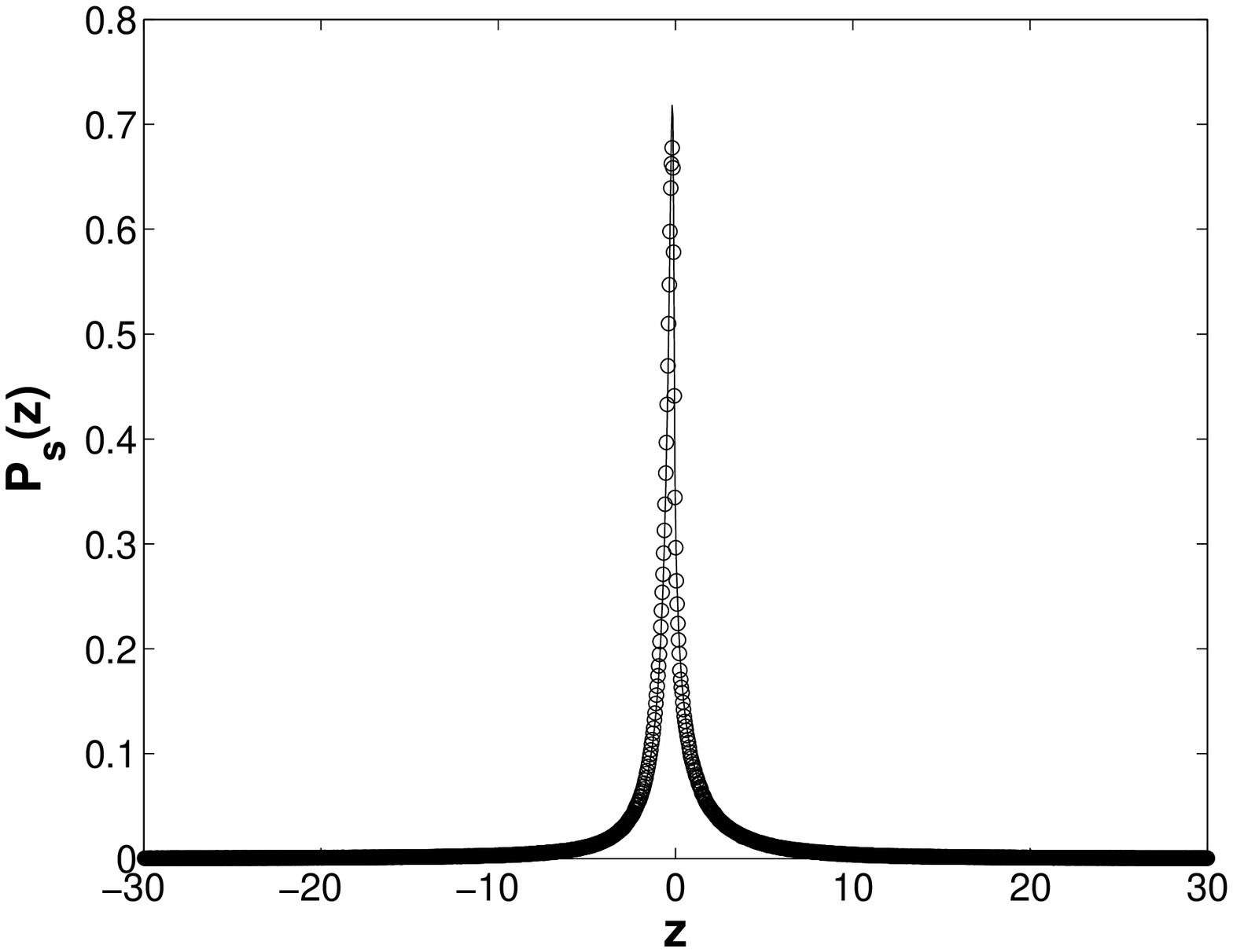}
\caption{\label{TransInt} Numerical simulations  of the 
 system (\ref{model2}) with $\gamma=1$, $k=1$ and $\Delta=3$ (top left
 panel) corresponding to an absorbing state and  for $\Delta=4.5$
 (top right panel), corresponding to an intermittent state. The lower
 panels compares   the  probability distribution for
 $z$  obtained by simulations (dots) 
  with  the  exact result~(\ref{Pstat2})  (full lines).}
\end{figure}
\end{center}

\begin{center}
\begin{figure}
\includegraphics[scale=0.4,clip]{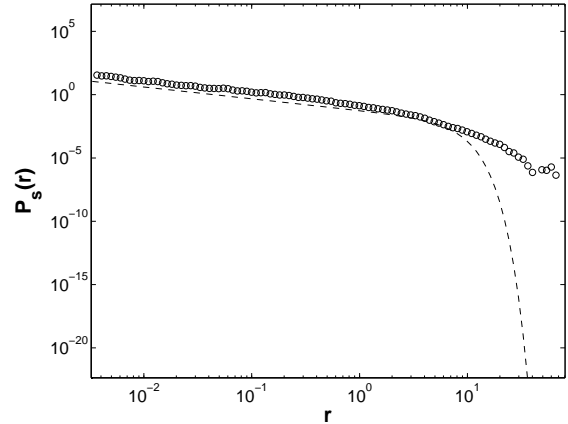}
\caption{\label{Pr} Probability density (in log-log scale) of the 
variable $r=\sqrt{x^2+\dot{x}^2}$ for $\gamma=1$, $\Delta = 4.2$ and
$k=1$.  The  dotted line  corresponds  to the  
approximation~(\ref{Psr}).}
\end{figure}
\end{center}

As shown in  figure \ref{TransInt}, the temporal series of the signal
 $x(t)$   exhibits  \lq\lq on-off intermittency\rq\rq above the
 instability threshold ({\it i.e.}, for a  positive Lyapunov exponent). 
  In other words,  the   amplitude  $x$  
 is vanishingly small for the 
   most of the time  but exhibits sudden  burst of
 activity. This behaviour has already been  observed  in chaotic systems
 \cite{Ott94,Sweet01} and in systems driven by multiplicative noise
 \cite{Japonais,Japonais2,Tresser,Leprovost05b,Aumaitre05}.  This
 intermittency can  also be identified in  the probability
 distribution density. In  figure \ref{Pr}, we show the probability
 density of the energy of the oscillator  defined as
 $r=\sqrt{x^2+\dot{x}^2}$. Near the origin, we observe that the probability
 distribution of $r$  is a power law   which is generic of an
 intermittent signal.  This   power-law distribution  for the energy
 of the oscillator can be derived  as follows.  From
 equation~(\ref{model2}), we deduce that $r$ satisfies \EQ \dot{r} =
 -(\gamma+\xi(t)) r \sin^2\theta  - \bigl(\gamma \sin^2\theta
 \cos^2\theta + k \cos^3\theta\sin\theta\bigr) r^3 \; , \nonumber \EN
 where $\theta = \arctan(\dot{x}/x) = \arctan(z)$. This equation,
 together with  (\ref{zpoint}), forms a system of stochastic
 equations.  A  Fokker-Planck equation for the joint density
 $P(r,\theta)$ or $P(r,z)$ can be written but the resulting   partial
 differential equation  seems unlikely to be exactly solvable.
 
 We therefore make an  approximation  along  the lines of
 \cite{Mallick03}  and assume that,  for small values of $r$, 
  the probability density is separable and can
 be written as  $P(r,\theta)=P_s(r) P_s(\theta)$. An  average over the
 angular variable yields  an independent equation for $P_s(r)$ that
 can be exactly  solved: \EQ
\label{Psr}
P_s(r) = \frac{1}{N_r}  r^{(B/A) -1}  
\exp\Bigl[-\frac{C r^2}{2A}\Bigr] \;
, \EN where the coefficients $A,B$ and $C$ can all be expressed as
mean values over the  variable  $z$ (or, equivalently, over the angular
variable $\theta$)~:
%  \EQA a &=& \frac{\Delta}{2} \langle
% \sin^4\theta \rangle = \frac{\Delta}{2} \langle \frac{z^4}{(1+z^2)^2}
% \rangle \\ \nonumber b &=& \langle \Delta \sin^2 \theta \cos^2 \theta
% - \gamma \sin^2 \theta \rangle \\ \nonumber &=&  \langle \Delta
% \frac{z^2}{(1+z^2)^2} - \gamma \frac{z^2}{(1+z^2)} \rangle \\
% \nonumber c &=& \langle \gamma \sin^2\theta \cos^2\theta + k
% \cos^3\theta \sin\theta \rangle \\ \nonumber &=& \langle \gamma
% \frac{z^2}{(1+z^2)^2} + k \frac{z}{(1+z^2)^2} \rangle \ENA
 \EQ 
 A =  \langle \frac{\Delta\,  z^4}{2(1+z^2)^2} \rangle  ,  \,   B
=   \langle  z^2 \frac{\Delta - \gamma(1+z^2)}{(1+z^2)^2} \rangle , \,
 C  =   \langle   \frac{\gamma z^2 + kz}{(1+z^2)^2}  \rangle  \, .\nonumber
  \EN 
 Multiplying equation~(\ref{FPstat})
by $z^2/(1+z^2)$ and integrating the result
 over $z$,  the following identity is
obtained~: \EQ  B=\langle z \rangle = \Lambda \; .   \EN  The
coefficient $A$ being a positive quantity, we  remark  that the
function~(\ref{Psr}) is normalizable if and only if  $B=\Lambda >0$.
This provides a simple derivation of the stability criterion that we
 have used~: when  $\Lambda < 0$, the fixed point
$x=0$ is stable (the distribution~(\ref{Psr}) being not normalizable,
the stationary distribution is the Dirac function at  0); when
$\Lambda >  0$, the fixed point
  is unstable and the  stationary distribution  is
extended.

\begin{center}
\begin{figure}
\includegraphics[scale=0.4,clip]{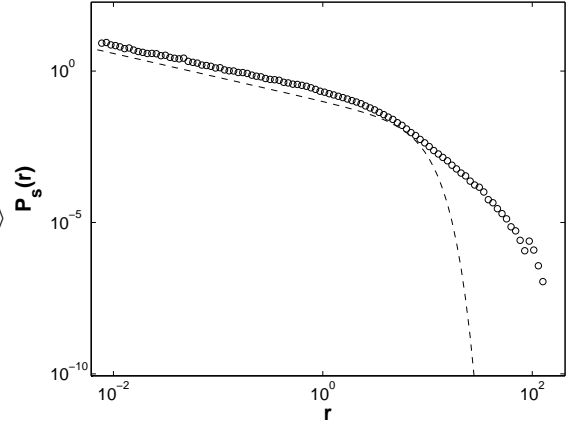}
\caption{\label{Pr2} Probability density (in log-log scale)
  of the variable $r=\sqrt{x^2+\dot{x}^2}$ for $\gamma=1$,
 $\Delta = 6$ and $k=1$.}
\end{figure}
\end{center}

The distribution (\ref{Psr}) is obviously a power law for  small
values of the variable $r$ which is an indication  for intermittency. In
figure \ref{Pr},  numerical results (open circles)  for
 the probability distribution of the
variable $r$  are  compared  with  the  analytical 
 formula~(\ref{Psr}) (dashed line);   the coefficients  $A$, $B$ and
$C$ have been calculated  using~(\ref{Pstat2}).
  The agreement between the two curves is excellent
as far as  the power-law behaviour for small  values of $r$ is
concerned.  For higher  values of $r$,  a  discrepancy  appears~:
the assumption that the stationary distribution is separable is no more
valid  and a specific analysis  for large $r$ is  needed
\cite{Mallick03}.  In figure~\ref{Pr2}, we plot  the probability
density of the energy for a greater value of the noise~: again the
small $r$ power law is very well described by equation~(\ref{Psr}).

\section{Effect of a colored noise}
\begin{center}
\begin{figure}
\includegraphics[scale=0.4,clip]{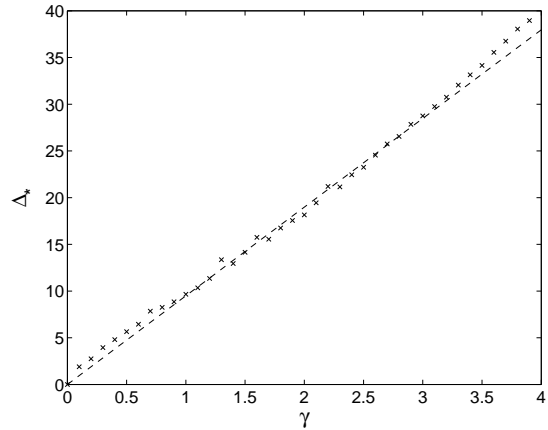}
\caption{\label{OU1} Critical amplitude $\Delta_{\star}$ of the
  noise versus $\gamma$ 
 for a time correlation $\tau=1$. The dotted line is a linear fit.}
\end{figure}
\end{center}
\begin{center}
\begin{figure}[h]
\includegraphics[scale=0.4,clip]{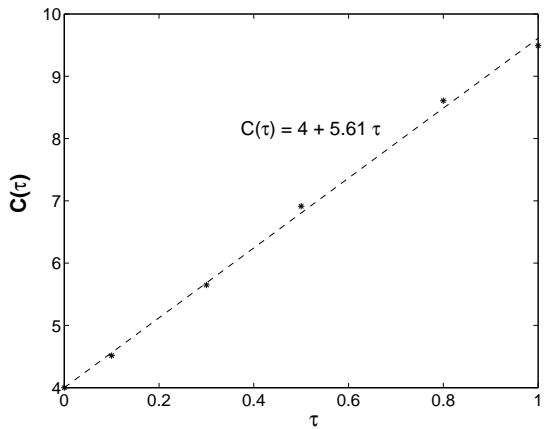}
\caption{\label{OU2} Evolution of the critical 
 slope as a function of the  correlation time.
  The dotted line is a linear fit.}
\end{figure}
\end{center}

 In order to  study the effect of the temporal correlations of the
 noise, we have performed  numerical simulations of equation
 (\ref{model2}),  taking  the noise to be  an Ornstein-Uhlenbeck process.
  In the stationary regime,  the correlation
 of the  Ornstein-Uhlenbeck process is  exponential~: \EQ \langle
 \xi(t) \xi(t') \rangle = \frac{\Delta}{2\tau} \exp\Bigl[-\frac{\vert
 t-t' \vert}{\tau}\Bigr] \; .  \EN  Using  the time series of the
 simulation results, we estimate  the Lyapunov exponent (by computing
 $\langle \dot{x}/x \rangle$)  and   identify the   bifurcation
 threshold.  In figure~\ref{OU1}, we  plot  the critical
 value of the noise intensity $\Delta$ for the onset of instability
  as a function of $\gamma$, the correlation time  $\tau$  being  equal
 to 1 (in dimensionless  units). We observe that  the critical curve  has
 the shape of  a straight line.  Performing  this analysis for
 different values of  $\tau$,  we have found  that the critical
   amplitude $\Delta_{\star}$ of the
  noise  is almost  linear with  
 $\gamma$; hence,  to a good approximation, the critical curve is
   given by   $\Delta_*(\gamma) = C(\tau) \gamma$.  In
 figure~\ref{OU2}, we plot  $C(\tau)$ as a function of $\tau$;  again
 the evolution is well described  by a straight line in the range  $ 0
 \le \tau \le 1$. These results indicate that the structure of the
 bifurcation diagram for the Ornstein-Uhlenbeck noise remains
 qualitatively the same as that  for white noise.

\section{Conclusion}
In this paper, we have studied the noise-induced  bifurcation  of a
nonlinear  oscillator with a fluctuating damping term.  We have shown
that the instability threshold of the nonlinear oscillator  can not be
determined from  a stability analysis of the  moments of  the
linearized problem; the system is indeed  more stable than the
analysis of moments  indicates.  This feature,  that  has already been
noticed in a variety of models  \cite{Graham82,Becker94,Mallick03},
can be explained by the existence of long tails in the probability
distribution  of the linearized system. These long tails
 have  a greater influence on higher moments  of $x$
 which are therefore  less and less stable.  However, these
long tails are suppressed for the nonlinear 
 system, which has, therefore,  a well defined stability threshold. 
  For the problem considered here,
 an exact calculation of the Lyapunov exponent has allowed
us to determine the  exact stability diagram of the system.  Besides, 
we have shown that  a  good quantitative description of the system  in
the vicinity of the bifurcation can be obtained if 
the  probability distribution is assumed
  to be separable  in energy and angle
variables.  This Ansatz  allows us  to calculate the power law
exponent of the stationary probability and to predict intermittency.
Our theoretical findings agree   with numerical simulations.
Finally, we have  studied numerically
  the effect of temporal correlation of the
noise  by coupling our dynamical system to an  Ornstein-Uhlenbeck
process,    and   have found   that the bifurcation
scenario  remains qualitatively the same.  For this latter problem,
exact analytical calculations  seem to be out of reach; some   quantitative
information may,  however, be obtained either by using one of the various 
Markovian approximations  for 
  colored noise, or by considering a special type of
random process, such as the dichotomic  Poisson noise which leads,  
 in the present case, to an exactly solvable problem.

\begin{acknowledgement}
We would like to thank Francois Petrelis and Stephan Fauve for
 interesting discussions and suggestions on this paper.
\end{acknowledgement}

\bibliographystyle{epj}

\bibliography{BiblioNick}

\end{document}